\documentclass[showpacs,pra,tightenlines,12pt]{revtex4-1}
\usepackage{amssymb,amsmath}
\usepackage{graphicx}
\usepackage{subfigure}

\def\ket#1{|#1 \rangle}
\def\bra#1{\langle #1 |}

\begin{document}

\title{SEARCH VIA QUANTUM WALKS WITH INTERMEDIATE MEASUREMENTS}

\author{EFRAIN BUKSMAN}
\email{buksman@ort.edu.uy}
\affiliation{Facultad de Ingenier\'ia Bernard Wand-Polaken, Universidad ORT Uruguay, Uruguay.}

\author{ANDR\'E L. FONSECA DE OLIVEIRA}
\email{fonseca@ort.edu.uy}
\affiliation{Facultad de Ingenier\'ia Bernard Wand-Polaken, Universidad ORT Uruguay, Uruguay.}

\author{JES\'US GARC\'IA L\'OPEZ DE LACALLE}
\email{jglopez@eui.upm.es}
\affiliation{Escuela Universitaria de Inform\'atica, Universidad Polit\'ecnica de Madrid, Spain.}

\date{\today }

\begin{abstract}
A modification of Tulsi's quantum search algorithm with intermediate measurements of the control is presented. In order to analyze the effect of measurements in quantum searches, a different choice of the angular parameter is used. The study is performed for several values of time lapses between measurements, finding close relationships between probabilities and correlations (Mutual Information and Cumulative Correlation Measure). The order of this modified algorithm is estimated, showing that for some 
time lapses the performance is improved, and became of order $O(N)$ (classical brute force search) when the measurement is taken in every step. The results indicate a possible way to analyze improvements to other quantum algorithms using one, or more, control qubits.
\end{abstract}


\maketitle

\section{Introduction}

Although there is no consensus on what are the quantum properties that cause the advantage of a quantum algorithm over classical counterparts, this advantage has been associated mainly with quantum correlation (concepts as entanglement, quantum discord, non-locality,  contextuality and others)\cite{Nest_2013}. While it is believed that some quantum correlation is necessary in quantum search, it is not exactly clear how  correlation is related with good results\cite{Cui_2010}. Correlation is modified by measurements and interaction with the environment, generally modeled as noise\cite{Nielsen_2000}, whose influence on quantum search algorithms has been extensively studied in last years\cite{Romanelli_2005,Salas_2008,Abal_2009,Srikanth_2010,Gawron_2012}. Furthermore, some research has been done to analyze the effect of projective partial measurements of the system\cite{Tulsi_2006,Maloyer_2007}.

It has been generally assumed that decoherence degrades the efficiency of a quantum 
algorithm. Nevertheless, in recent years some studies have shown that low levels of noise 
can improve certain algorithms\cite{Kendon_2007,Venegas_Andraca_2012}. Would this be the 
case of projective partial measurements in quantum searches? In this article we show that 
the effect of intermediate partial measurements during the state evolution can improve 
the performance of some search algorithms, i.e. controlled quantum walk search 
algorithms. In addition, we show there is a close relationships between the search results and correlations.

The proposed quantum search is an intermediate case between two known methods: unitary algorithms that starts with separable states\cite{Portugal_2013}, where the correlation increases during the evolution; and Measurement Based Quantum Computation (MBQC)\cite{Raussendorf_2003}, where the initial states are of high correlation (cluster states), and measurements are taken to achieve the desired results.

Tulsi's algorithm\cite{Tulsi_2008} is an special case of the abstract search 
algorithm based on a discrete quantum walk, that uses a control qubit in addition to the usual coin. This quantum walk depends on an angular parameter $\delta$, where the optimum value ensures an order of $O(\sqrt{N \log N})$, considering a $\sqrt{N} \times \sqrt{N}$ search grid. 

In this article a modified instance of Tulsi's algorithm, in a Hilbert space $\mathcal{H} = (\mathbb{C}^2)^{\otimes m}$, is used with the main purpose of analyze the influence of partial intermediate measurement in quantum algorithms. With this aim, and considering that Tulsi's algorithm is optimal, a different choice of the angular parameter $\delta$ is made. Based on numerical evidence, it has been found that a unitary algorithm can be improved by performing partial projective measurements with a convenient choice of time lapses between unitary evolutions. Relationship among the time between intermediate measurements, the correlation, and the order of the algorithm is discussed. 

The paper is organized as follows. In order to introduce the notation and operators used, Tulsi's algorithm is briefly explained in Section 2. Modified Tulsi's algorithm with intermediate partial measurements is presented in Section 3. Section 4 deals with the  relation between probabilities and correlations. In Section 5, the order of the presented algorithm is estimated depending on time lapse between measurements. Finally, some conclusions and proposals for future works are commented in Section 6. 

\section{Unitary Tulsi algorithm}

Tulsi's quantum walk based algorithm searches an unique item out of $N$ items arranged
on a two-dimensional ($\sqrt{N} \times \sqrt{N}$) grid. This position space is
represented by an n-qubit quantum state ($N = 2^{n}$), and it is initialized with a state formed by an uniform superposition of the canonical basis, similar to Grover\cite
{Grover_1996} and AKR\cite{AKR_2005} quantum algorithms. In addition to the position, the state has a two qubits coin state and a control qubit (initialized as indicated in (\ref{estado_inicial})).

The Tulsi's algorithm uses two operators: the oracle and the the conditional walk. The conditioned reflection operator, called oracle, is given by\footnote{In this article the subscripts $ctr$, $c$ and $p$, indicate the control, coin and position subspaces, respectively.},
\begin{eqnarray}
O = I - 2 \, \ket{\delta_{ctr},u_c,t}\bra{\delta_{ctr},u_c,t}
\label{Oraculo}
\end{eqnarray}
where
\begin{itemize}
\item the control qubit $\ket{\delta_{ctr}} = -\sin \delta \ket{0}+\cos \delta \ket{1}$,
depends on the $\delta$ parameter,
\item the coin state $\ket{u_c}$ is a balanced superposition of two qubits in the coin basis $\{\ket{i_c}\}=\{\leftarrow,\rightarrow,\downarrow,\uparrow \}$
\begin{eqnarray}
\ket{u_c}=\frac{1}{2}\sum_{i_c=0}^{3} \ket{i_c},
\label{u_c}
\end{eqnarray}
\item and the target state $\ket{t}$ is any unknown state of the position subspace.
\end{itemize}

The conditional walk operator performs a walk depending on the value of the control qubit, as showed in the following equation
\begin{eqnarray}
W =\ket{1}\bra{1} \otimes (S \cdot C)  -\ket{0}\bra{0}\otimes I_{c,p}
\label{W}
\end{eqnarray}
where
\begin{itemize}
\item $S$ is the shift operator
\begin{eqnarray}
S=\sum_{x,y} \ket{\leftarrow ,x+1,y}\bra{\rightarrow ,x,y}+\ket{\downarrow   ,x,y+1}\bra{\uparrow, x,y}+\nonumber \\
\ket{\rightarrow ,x-1,y}\bra{\leftarrow, x,y}+\ket{\uparrow, x,y-1}\bra{\downarrow, x,y} \nonumber\\
\end{eqnarray}
being the state $\ket{x,y}$ related to the $x,y$ position on the grid,
\item $C$ operates only in the coin subspace\cite{AKR_2005}
\begin{eqnarray}
C_0= -I_c +2 \ket{u_c}\bra{u_c}
\end{eqnarray}
where $ C= C_0 \otimes I_p$.
\end{itemize}

\begin{figure}[ht]
\centering
\includegraphics[width=0.6\textwidth,keepaspectratio=true]{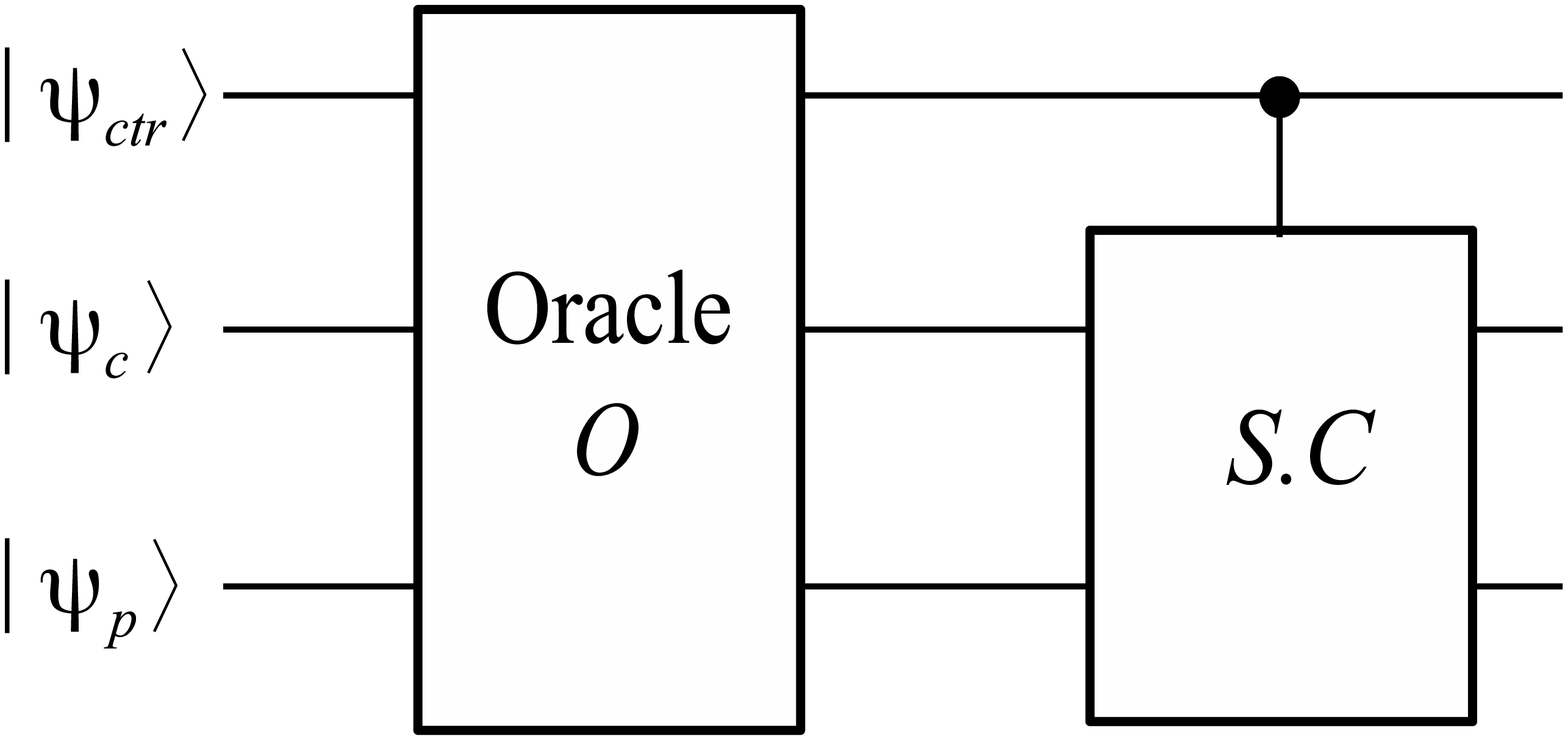}
\caption{Tulsi algorithm}
\label{Fig_Alg_Tulsi}
\end{figure}

As proved in \cite{Tulsi_2008}, considering an initial state $\ket{\psi_0}$
\begin{eqnarray}
\label{estado_inicial}
\ket{\psi_0}=\ket{1,u_c,u_p}= \ket{1} \otimes \ket{u_c}  \otimes \ket{u_p} = \ket{1} \otimes \ket{u_c} \otimes \frac{1}{\sqrt{N}}\sum_{i_p=0}^{N-1} \ket{i_p},
\end{eqnarray}
and applying $T_{\delta}=[ \frac{\pi}{4} \sqrt{N  (\log{N}+\tan{\delta}^2/4 )} ]$ times (steps) the operator $U=W.O$ (Fig. \ref{Fig_Alg_Tulsi}), the evolving state approaches the target state. The main result is that the overlap between the final state and the target state depends on $\delta$ and $N$.
By choosing parameter $\delta$ as $\cos(\delta)=1/\log{N}$, the target probability becomes independent of $N$, and the algorithm has an order $O(\sqrt{N \log{N}})$.

\begin{figure}[ht]
\centering
\includegraphics[width=0.75\textwidth,keepaspectratio=true]{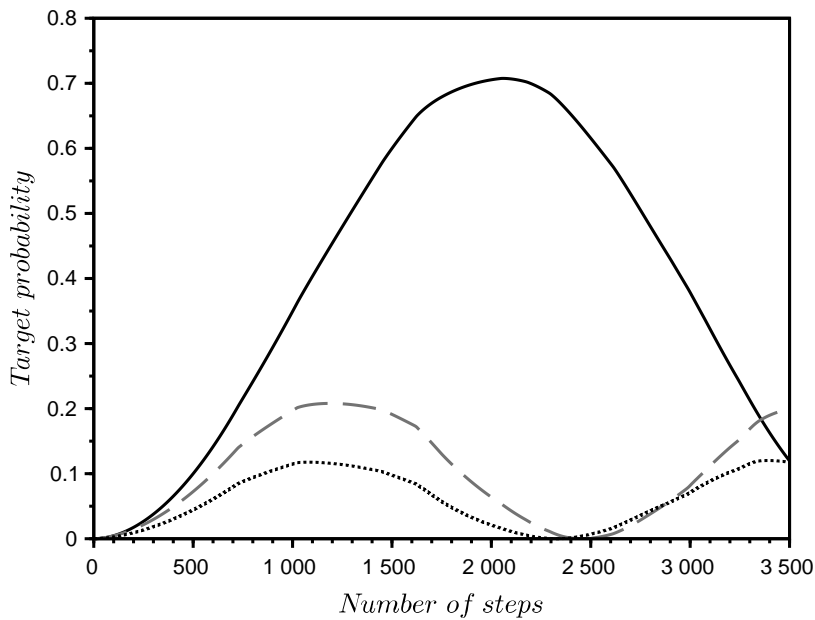}
\caption{Unitary evolution of Tulsi algorithm with $N=2^{18}$ and: $\delta=0$ (AKR with dots), $\delta=\pi/4$ (dash line) and $\delta=\arccos({\sqrt{1/\log{N}}})$ (solid line).}
\label{Tulsi_Jae_AKR}
\end{figure}

For a $2^{9} \times 2^{9}$ grid ($N=2^{18}$), and three angular parameters choices ($\delta=0$, $\delta=\pi/4$ and the Tulsi's parameter $\delta=\arccos \left( 1/{\sqrt{\log{N}}} \, \right)$), the resulting probabilities are shown in Figure (\ref{Tulsi_Jae_AKR}).

\section{Intermediate measurements algorithm}
\label{sec_IMA}
\begin{figure}[ht]
\centering
\includegraphics[width=0.6\textwidth,keepaspectratio=true]{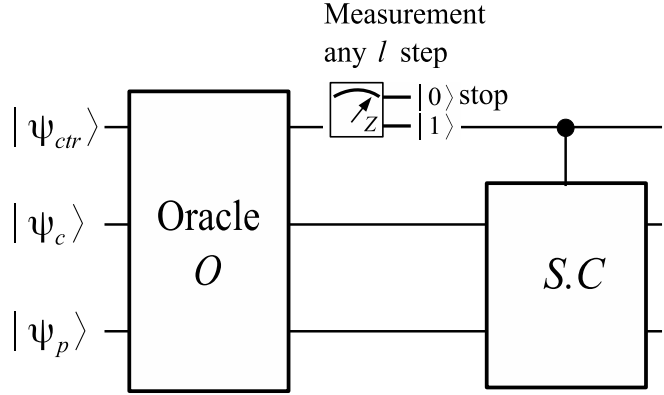}
\vspace*{8pt}
\caption{Intermediate measurements algorithm: $IMA$}
\label{Fig_Algoritmo_IMA}
\end{figure}

Unlike Tulsi's unitary algorithm, our proposal (Figure (\ref{Fig_Algoritmo_IMA}))  consists in taking projectives measurements of the control qubit, at time-lapses $l$, between unitary evolutions. The aim is to understand how intermediate partial measurements affect the target probability, and its relationship with correlation.

After any measurement is performed, the obtained information allows to stop the algorithm if the result is $0$, since this control value is exclusively related to the target state position. This can be seen as the result of applying the sequence of operators (\ref{Oraculo}) and (\ref{W}) to a  basis state $\ket{i_{ctr},i_c,i_p}$. In the case where $\ket{i_p}$ is any position basis state different from the target position we have
\begin{eqnarray} 
\ket{1,i_c,i_p} \xrightarrow{O} \ket{1,i_c,i_p} \xrightarrow{W}
\left\{  
\begin{array}{ccc} 
\ket{1,{i_c}',{i_p}'} \\
\ket{1,{i_c}',t} \\
\end{array} 
\right.
\end{eqnarray} 
Similarly, for a basis state with a target position we obtain
\begin{eqnarray} 
\ket{1,i_c,t} \xrightarrow{O}
\left\{
\begin{array}{ccc} 
 \ket{1,{i_c}',t} \\
 \ket{0,{i_c}',t} \\
\end{array} 
\right.
\xrightarrow{W} 
\left\{
\begin{array}{ccc} 
\ket{1,{i_c}'',i_p} \\
\ket{0,{i_c}',t} \\
\end{array} 
\right.
\nonumber
\end{eqnarray}
\begin{eqnarray} 
\ket{0,{i_c},t}  \xrightarrow{O} 
\left\{
\begin{array}{ccc} 
\ket{0,{i_c}',t} \\
\ket{1,{i_c}',t} \\
\end{array} 
\right.
\xrightarrow{W}
\left\{
\begin{array}{ccc} 
\ket{0,{i_c}',t} \\
\ket{1,{i_c}'',i_p} \\
\end{array} 
\right.
\end{eqnarray}
Therefore, it is concluded that starting with state given by (\ref{estado_inicial}),
no state of the type $\ket{0,i_c,i_p}$ could appear during the evolution if  $\ket{i_p}$ is different from the target state $\ket{t}$. 

\vspace*{14pt}
\noindent
{\textit{Intermediate measurements algorithm (IMA)}}

The controlled quantum walk with intermediate measurements has the following algorithm:
\begin{enumerate}
\item The system is initialized at state $\ket{\psi_0}$ (Eq. \ref{estado_inicial}).
\item Apply $l$ times the $U=W O$ operator.
\item Measure the control qubit.
\item If the measurement result is $0$ stop algorithm: the target is found.
\item Otherwise, return to step 2 until a maximum of $k_{max}$ total steps are reached.
\item After $k_{max}$ total steps, check if the position is the target state. If this is not the case, start over (from step 1, at state $\ket{\psi_0}$).
\end{enumerate}
Similar to Tulsi's algorithm, in this article we will use a value of $k_{max}=(\pi/4)(\sqrt{N \log N })$. This value is approximately the optimal $k_{max}$ step in which the algorithm should stop, as shown in section 4.

The cumulative target probability $P_c$, for any time lapse $l$ ($l=1,2,\dots, k_{max}$)
is given by                                                         
\begin{eqnarray} 
P_c(k,l) & = & P_0^l +P_1^l P_0^{2 l} + P_1^l P_1^{2 l}P_0^{3 l} + \dots + P_1^l \dots P_1^{\lfloor {k/l}\rfloor l} P_t(k,l) \label{Prob_acumulada} \\
& = & P_0^l + (1 - P_0^l)P_0^{2 l} + \dots + (1 - P_0^l)(1 - P_0^{2l})\dots(1 - P_0^{\lfloor {k/l}\rfloor l})P_t(k,l) \nonumber \\
& = & P_0^l + (1 - P_0^l) \left[ P_0^{2 l} + (1 - P_0^{2 l}) \left[ \dots \left[  P_0^{\lfloor {k/l}\rfloor l} + (1- P_0^{\lfloor {k/l}\rfloor l}) P_t(k,l) \right] \right] \right], \nonumber
\end{eqnarray}
where
\begin{itemize}
\item $P_t(k,l)=| \langle {\psi} | \left( I_{ctr} \otimes I_c \otimes \ket{t} 
\bra{t} \right) | {\psi} \rangle |^2 $ is the target probability at a given step $k$,
\item $P_0^{m l}$ and $P_1^{m l}$ are the corresponding probabilities of values $0$ and 
$1$ at a step multiple of $l$,
\item and $\lfloor {k/l}\rfloor$ is the integer part of $k/l$.
\end{itemize}

As each bracket in (\ref{Prob_acumulada}) has the form $x + (1-x)a$ ($0 \leq x,a \leq 1$), $P_c(k,l)$ is always in $[0,1]$. 

Given the non-unitary nature of the algorithm, rather than quantum amplification\cite{Benioff_2002}, classical amplification is used to obtain a search probability of order one. In this article, two cases of interest are studied: $\delta=\pi/4$, ($IMA_{\pi/4}$) and $\delta=\arccos \left( 1 / \sqrt{\log{N}} \right)$, ($IMA_T$). For $\delta=0$ the algorithm is identical to AKR algorithm, and the control does not affect the search\footnote{The examples were performed using QuantumLab\cite{Oliveira_2007}, a quantum simulator toolbox for Scilab.}.

\section{$IMA$: probabilities and correlations}
\label{sec_IMA_prob_cor}

In contrast to unitary algorithm, probabilities change drastically when intermediate partial measurements are performed. In this section, the target probability $P_t$ and the cumulative target probability $P_c$ are calculated and compared with correlation. Both probabilities depend on the measurement time lapse $l$. In the next numerical experiments different $l$ were chosen as a function of $N$,
\begin{equation}
l= \sqrt{N}/2^m, \quad m=1,2,3,\ldots.
\end{equation}
For comparison with the classical case a constant value of $l = 1$ is used in some experiments.

\begin{figure}[ht!]
\centering
\includegraphics[width=0.75\textwidth,keepaspectratio=true]{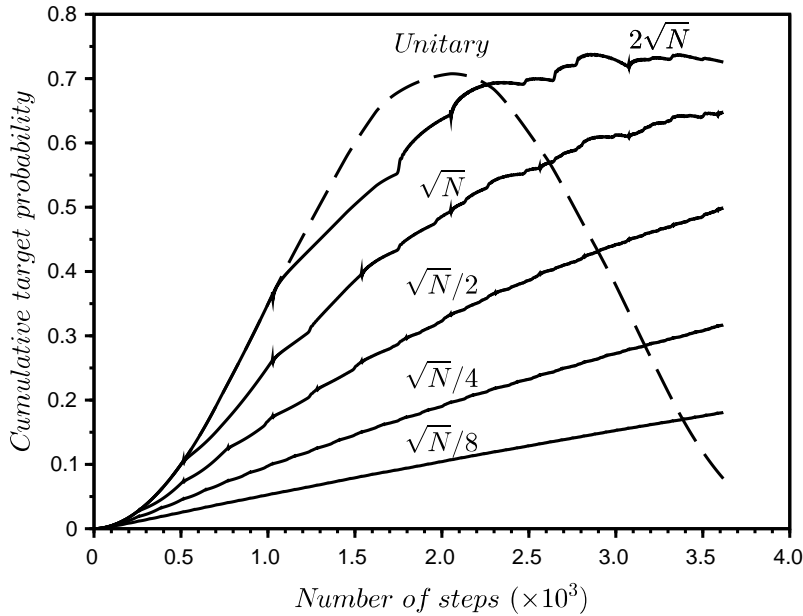}
\vspace*{8pt}
\caption{Cumulative target probability in $IMA_T$ algorithm ($\delta=\arccos \left( 1 / \sqrt{\log{N}} \right)$) for $N=2^{18}$ and $l=2\sqrt{N},\sqrt{N},\sqrt{N}/2, \sqrt{N}/4, \sqrt{N}/8$ and unitary case(dash line) }
\label{Probe_Tulsi}
\end{figure}

In the $IMA_T$ algorithm (original $\delta$ Tulsi's parameter), intermediate measurements 
always reduces the cumulative probability (for any $l$), as can be seen in Figure (\ref
{Probe_Tulsi}). Considering this, the unitary Tulsi's algorithm is optimal, i.e. it cannot be further improved by measurements.

On the other hand, when a different $\delta$ is considered, as in the $IMA_{\pi/4}$ 
algorithm, the cumulative probability can be improved, depending on time lapse $l$ used. In order to get symmetric operators (\ref{Oraculo}), in the following numerical studies a $\delta=\pi/4$ is chosen.

\begin{figure}[ht!]
\centering
\includegraphics[width=0.75\textwidth,keepaspectratio=true]{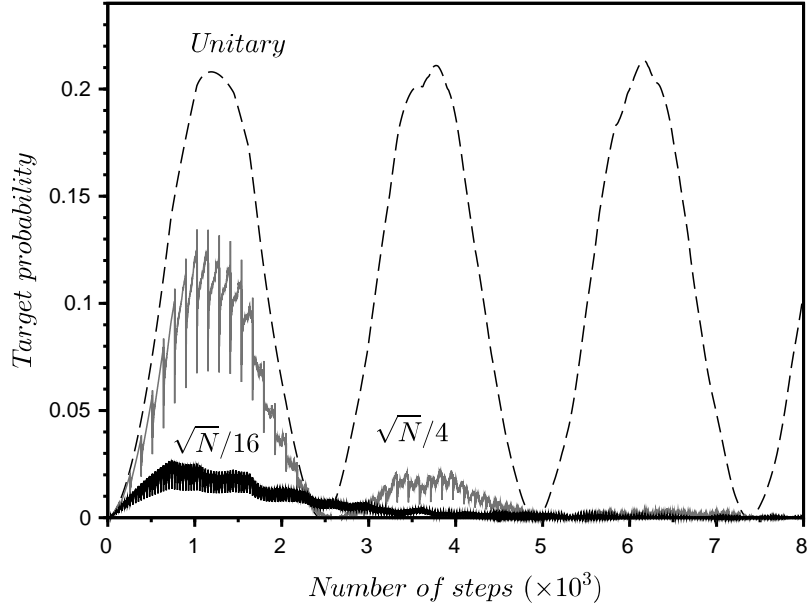}
\caption{Target probability on $IMA_{\pi/4}$ algorithm ($\delta=\pi/4$) with $N=2^{18}$. For 
time lapses: $l=\sqrt{N}/4$ (gray), $l=\sqrt{N}/16$ (black), and the unitary evolution (dashed line).}
\label{Pt_9q}
\end{figure}

\begin{figure}[ht!]
\centering
\includegraphics[width=0.75\textwidth,keepaspectratio=true]{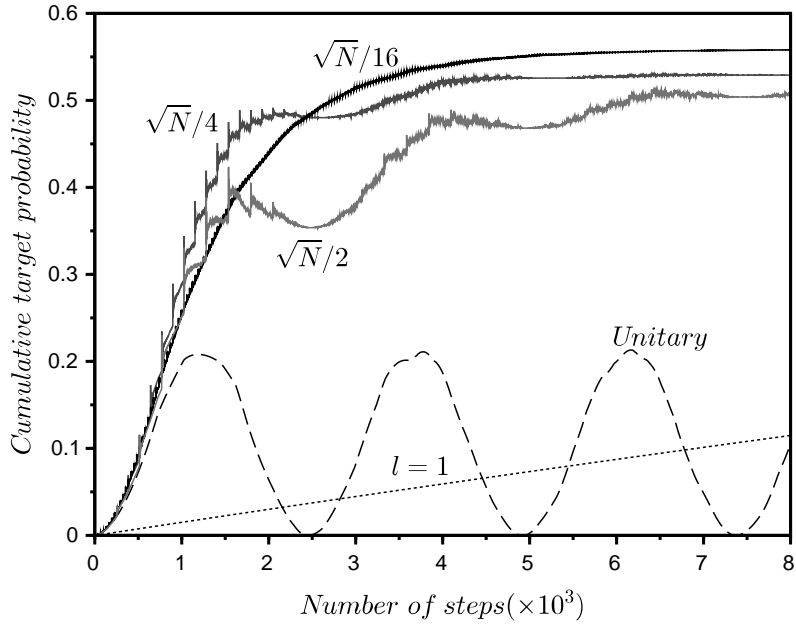}
\caption{Cumulative target probability on $IMA_{\pi/4}$ algorithm ($\delta=\pi/4$) with $N=2^{18}$. For 
time lapses: $l=\sqrt{N}/2$ (light gray), $l=\sqrt{N}/4$ (gray), $l=\sqrt{N}/16$ (black),  $l=1$ (dotted line) and the unitary evolution (dashed line).}
\label{Pc_9q}
\end{figure}

The target probability in the unitary walk (for any $\delta$) can be approximated by
a harmonic oscillation\cite{Tulsi_2008}. As expected, when a measurement is performed on the control qubit, the state loses coherence due to the fact it is strongly correlated with the rest of the state. This fact, combined with the effect of stopping the algorithm for a zero measurement at the control qubit, causes the target probability $P_t$ to go near zero, similarly to energy in a damped harmonic oscillator (see Fig. (\ref{Pt_9q})).  
The latter makes the probability $P_c$ tend to constant value for long times, as shown in Fig. (\ref{Pc_9q}). 

\subsection{Correlations}
\label{subsec_IMA_prob_cor_correlations}

In recent years some authors studied the behavior between probabilities and state correlations in quantum search algorithms\cite{Cui_2010}. In this section we analyze 
the relation of $P_c$ and $P_t$ with some correlation measures, as the
bipartite Mutual Information\cite{Vedral_2002} ($MI$) and the multipartite cumulative 
correlation\cite{Fonseca_2014} ($CCM$). A major advantage of these correlations is that they do not need nonlinear optimization methods. This is important because of the large number of states used in our quantum search.

In $IMA$ the state space have natural partitions, composed by the control, the coin and the position. The bipartite correlation that comes from partitioning the space in 
$\mathcal{H}_{control} \otimes \mathcal{H}_{coin+position}$, called $MI_{ctr\otimes(c,p)}$, is 
given by 
\begin{eqnarray} 
MI_{ctr\otimes(c,p)}(\rho) = -S(\rho)+S(\rho_{ctr})+S(\rho_{c,p})
\label{Corre_bipartita_ct-(c,p)}
\end{eqnarray}
where $\rho$ is the total density matrix of the state, $\rho_{ctr}$ is the reduced matrix of the control qubit, $\rho_{c,p}$ is the reduced matrix of the coin-position subspace, and $S$ the Von Neumann entropy.  

Other bipartite correlations are: $MI_{(ctr,c)\otimes p}$ and $MI_{(p,ctr)\otimes c}$. In these cases the state $\rho$ is always pure, so $S(\rho)=0$. Therefore, the 
distinction between classical and quantum parts is irrelevant, and can be considered a measure of entanglement\cite{Zhang_2012}. 

We consider also others bipartite correlations of mixed states, as the $MI$ between the 
coin and the position ($MI_{c\otimes p}$), the control and the coin ($MI_{ctr\otimes c}$), and the control and the position ($MI_{ctr\otimes p}$). For example, the 
$MI_{c\otimes p}$ is given by  
\begin{eqnarray} 
MI_{c\otimes p}(\rho) = -S(\rho_{c,p})+S(\rho_{c})+S(\rho_{p}),
\label{MI_c_p}
\end{eqnarray}
were, in this case, $S(\rho_{c,p}) \neq 0$.

\subsection{$P_t$ versus correlations}
\label{subsec_IMA_prob_cor_Pt}

\begin{figure}[ht!]
\centering
\includegraphics[width=0.75\textwidth,keepaspectratio=true]{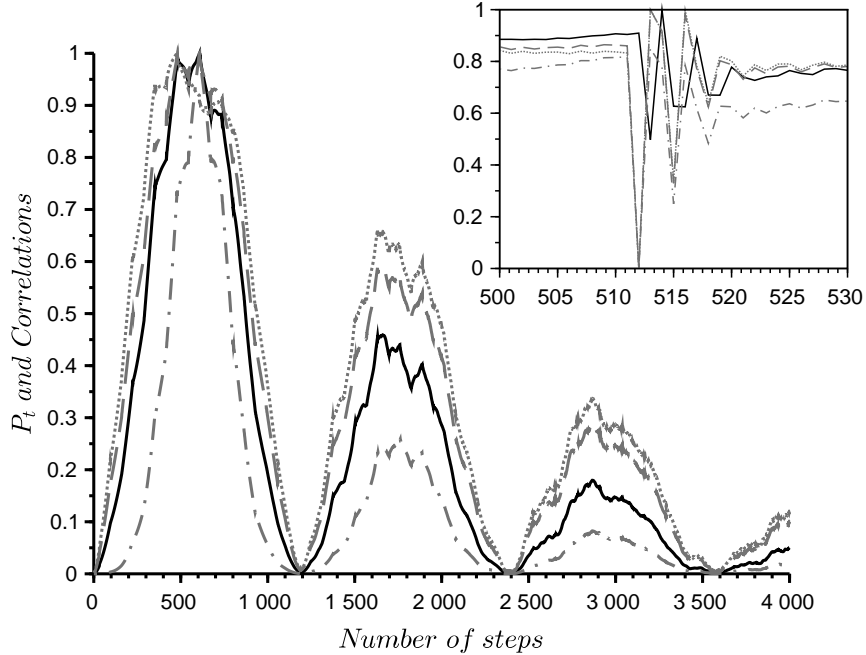} 
\caption{Target probability $P_t$(black solid) compared with correlations: 
$MI_{ctr \otimes (c,p)}$(dashed gray), $MI_{ctr \otimes p}$(dotted gray) and $MI_{ctr \otimes c}$(dash dotted gray), for $N=2^{16}$ and $l= \sqrt{N}/2$, all normalized and smoothed. In detail, original curves without smoothing.}
\label{Pt_and_Corre}
\end{figure}

Due to the explicit correlation imposed by the oracle (\ref{Oraculo}), the 
target probability $P_t$ shows a similar behavior as the correlations that isolate the control qubit: $MI_{ctr\otimes(c,p)}$, $MI_{ctr\otimes p}$ and $MI_{ctr\otimes c}$ (see Fig. (\ref{Pt_and_Corre})). These correlations must be zero 
immediately after the measurement of the control qubit as seen in the detail in Fig. (\ref
{Pt_and_Corre}). The fluctuations that appear after the measurement, both in 
the probability and the correlation, are due to secondary waves commonly related to 
quantum walks\cite{Knight_2003}. As fluctuations become smaller with increasing $N$, and with the intention of assessing the average behavior, the curves are smoothed by taking a suitable average for each case.

\begin{figure}[ht!]
\centering
\includegraphics[width=0.75\textwidth,keepaspectratio=true]{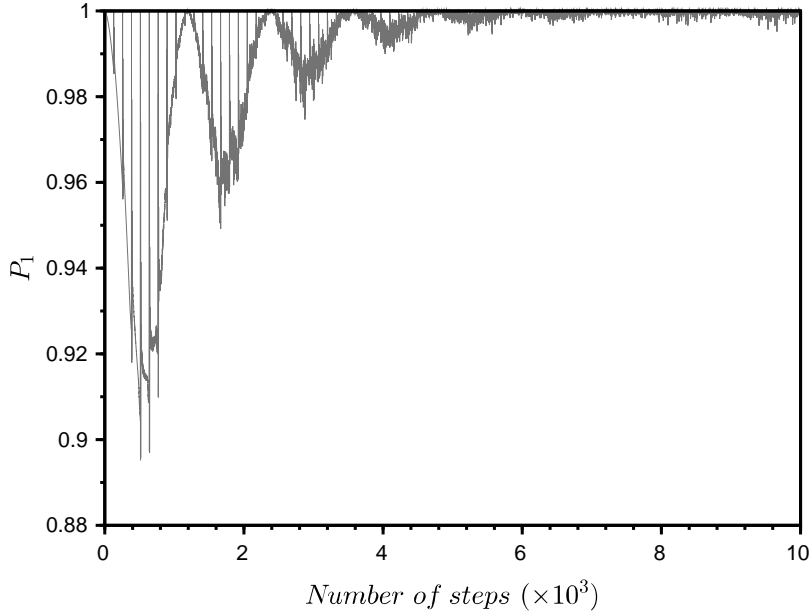} 
\caption{Probability of the control qubit being in state $\ket{1}$. Case for $N = 2^{16}$ and $l = \sqrt{N}/2$.}
\label{ProbDe1}
\end{figure}

Figure (\ref{ProbDe1}) shows the evolution of the probability of obtaining the state $\ket{1}$ in the control qubit. As can be observed, with the growth of the number of steps, the control qubit $\rho_{(ctr)}$ tends to be very near to a pure $\ket{1}$ state, and $S(\rho_{(ctr)}) \approx 0$.

Considering equation (\ref{Corre_bipartita_ct-(c,p)}) and that $\rho$ is a pure state,
\begin{equation}
MI_{ctr \otimes (c,p)} = 2 S \left( \rho_{ctr} \right) \approx 0.
\label{eqn_MI_ctr1}
\end{equation}

In the case of the others correlations, we have
\begin{eqnarray}
MI_{ctr \otimes p} & = & - S \left( \rho_{ctr, p} \right) + S \left( \rho_{ctr} \right) + S \left( \rho_{p} \right), \nonumber \\
 & \approx & - S \left( \rho_{ctr, p} \right) + S \left( \rho_{p} \right), \nonumber \\
 & \approx & 0. 
\label{eqn_MI_ctr2}
\end{eqnarray}
Similarly, $MI_{ctr \otimes c} \approx 0$. 

For a constant $l=1$, the bipartite correlations $MI_{ctr\otimes(c,p)}$, $MI_{ctr\otimes p}$ and $MI_{ctr\otimes c}$, are always zero, and correspondingly $P_t$
is very low and $P_c$ is almost linear (Figure (\ref{Pc_9q})). 

\subsection{$P_c$ versus correlations}
\label{subsec_IMA_prob_cor_Pc}

\begin{figure}[ht!]
\centering
\subfigure[Correlations $MI_{(p,ctr) \otimes c}$ (dark gray), $MI_{(ctr,c) \otimes }$ (black) and $MI_{c \otimes p}$ (light gray).  As the number of steps grows all correlations tends to be similar.]{
\label{PC_Correlations1}
\includegraphics[width=0.45\textwidth]{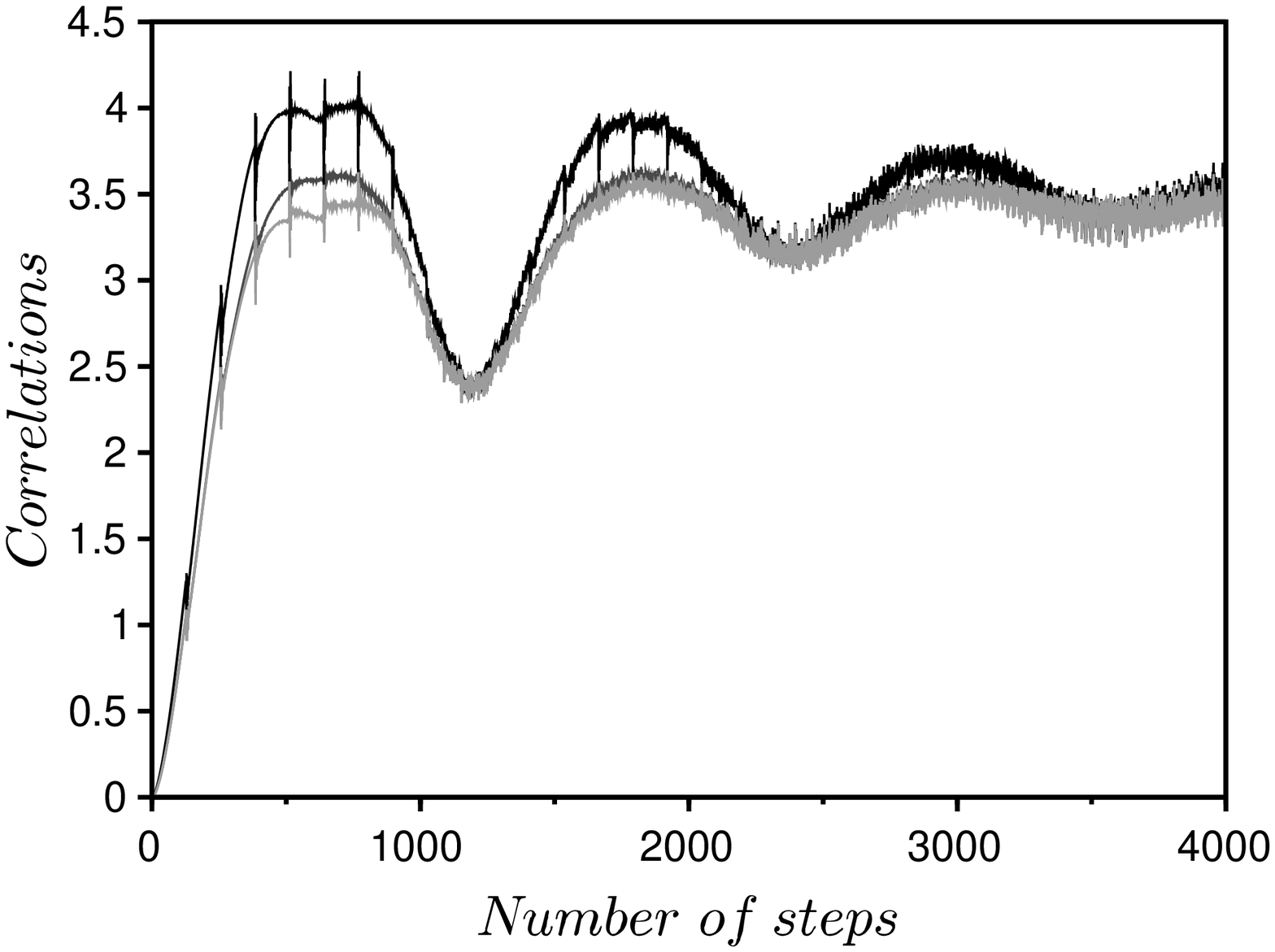}}
\hspace{0.05\textwidth}
\subfigure[Cumulative target probability $P_c$]{
\label{PC_Correlations2} 
\includegraphics[width=0.45\textwidth]{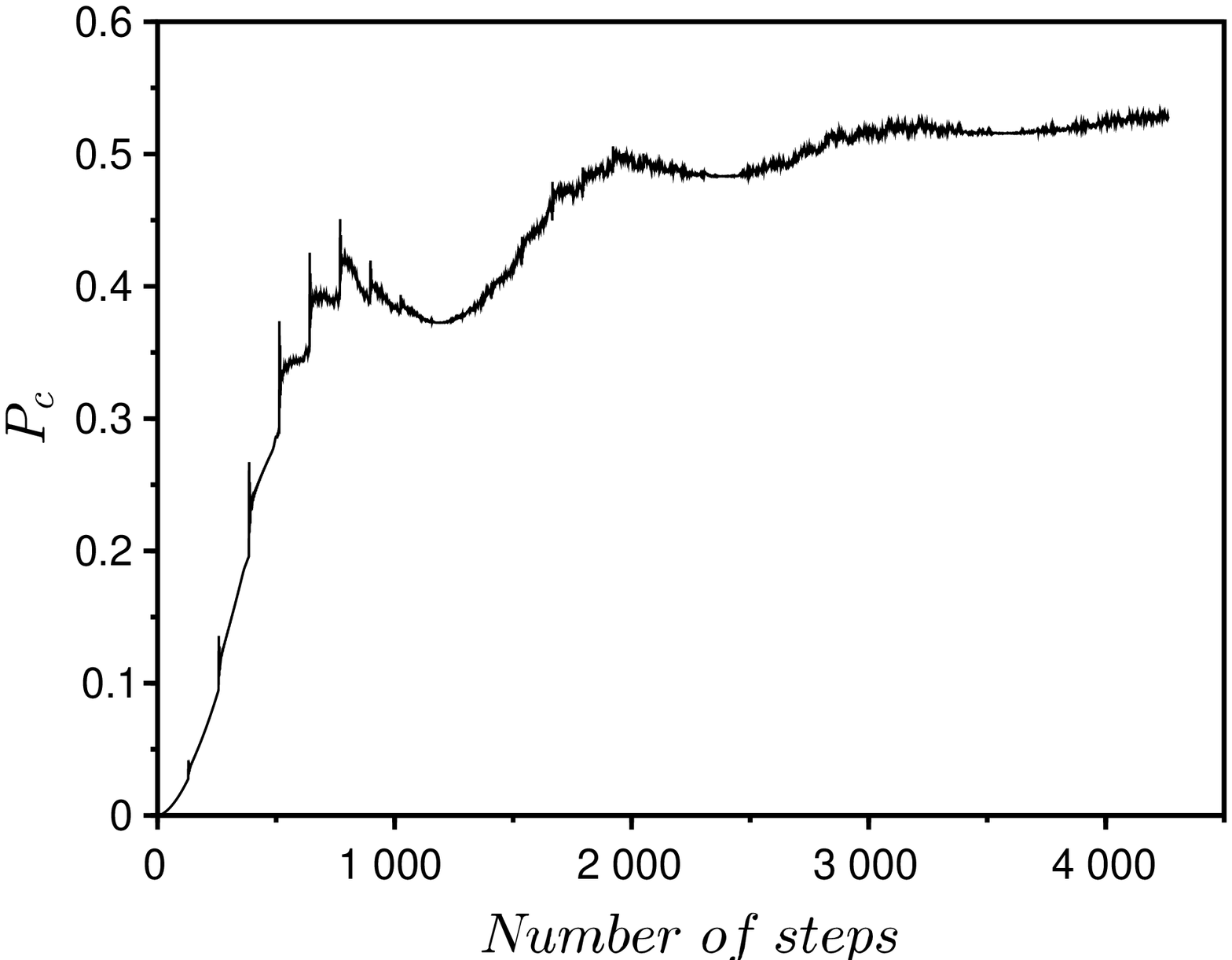}}
\caption{Cumulative target probability $P_c$ compared with correlations, for $N=2^{16}$ and $l= \sqrt{N}/2$.}
\label{PC_Corre} 
\end{figure}

On the other hand, the cumulative probability $P_c$ has a similar behavior compared to the correlations $MI_{(p,ctr) \otimes c}$, $MI_{(ctr,c) \otimes p}$ and $MI_{c \otimes p}$. These correlations oscillate, until they stabilize ($\pm 5\%$) approximated in $5  k_{max}$ steps, as can be seen in Fig. (\ref{PC_Corre}). This behavior is expected, since with the increased number of steps, the influence of the control qubit becomes negligible (\ref{eqn_MI_ctr1}). Therefore, for a large number of steps
\begin{eqnarray}
MI_{(p,ctr) \otimes c} & = & - S \left( \rho \right) + S \left( \rho_{p, ctr} \right) + S \left( \rho_{c} \right) \approx  S \left( \rho_{p} \right) + S \left( \rho_{c} \right), \\
MI_{(ctr,c) \otimes p} & = & - S \left( \rho \right) + S \left( \rho_{ctr, c} \right) + S \left( \rho_{p} \right) \approx  S \left( \rho_{p} \right) + S \left( \rho_{c} \right), \\
MI_{c \otimes p} & = & - S \left( \rho_{c, p} \right) + S \left( \rho_{p} \right) + S \left( \rho_{c} \right) = - S \left( \rho_{ctr} \right) + S \left( \rho_{p} \right) + S \left( \rho_{c} \right) \nonumber \\
& \approx & S \left( \rho_{p} \right) + S \left( \rho_{c} \right).
\label{eqn_MI_ctr3}
\end{eqnarray}

\begin{figure}[ht]
\centering
\includegraphics[width=0.75\textwidth,keepaspectratio=true]{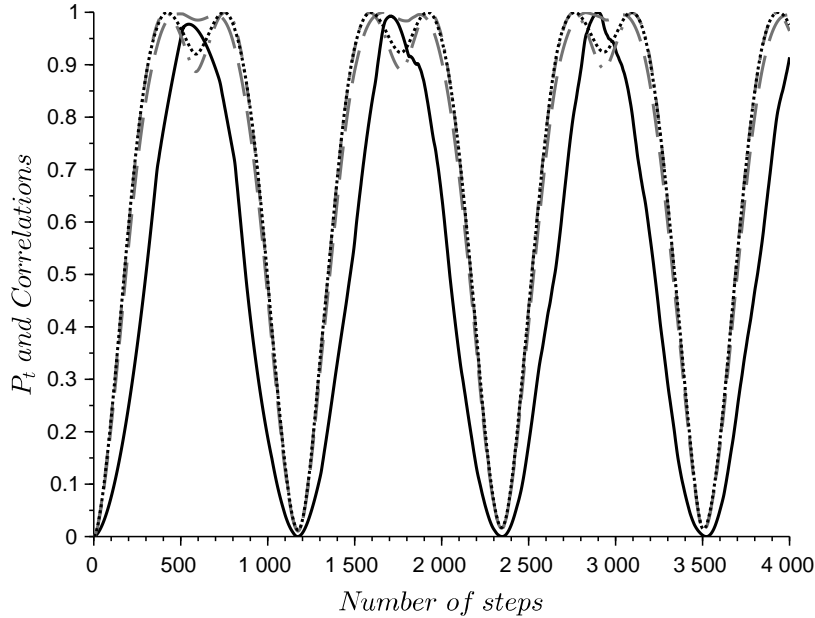}
\caption{Target probability $P_t$ compared with Correlations: $MI_{(p,ctr) \otimes c}$(dashed gray), $MI_{(ctr,c) \otimes p}$(dash dotted gray) and $MI_{c \otimes p}$(dotted black), for $N=2^{16}$ for unitary evolution all normalized.}
\label{MI_ctc_p_and_Prob}
\end{figure}

For unitary evolutions these correlations have also similar behaviors. They have a relative minimum where the target probability reaches a maximum, as shown in Fig. (\ref{MI_ctc_p_and_Prob}). This effect is caused by the convergence of the position subspace towards the target position, becoming less correlated with the rest of the state around the step of maximum probability. 

The same phenomena has been observed in Grover's search algorithm\cite{Cui_2010}. It has been found that Concurrence\cite{Wootters_2001} works as an indicator for the increasing rate of probability. Unlike the unitary $IMA_{\pi/4}$, in Grover's algorithm the target probability increases to values near to one, and at the same time the correlation decreases approximately to zero\cite{Grover_1996}.   

\subsection{Total steps versus correlations}
\label{subsec_IMA_Ts_cor}

In order to obtain an arbitrary search probability $P$, i. e. near $1$, amplification needs to be applied. As mentioned in section \ref{sec_IMA}, due to intermediate measurements, we apply classical amplification. Given an experiment with probability of success $P_0$, the total number of independent repetitions $R$ needed to obtain an arbitrary probability $P$ can be calculated as (geometric distribution)
\begin{eqnarray}
P & = & 1 - (1-P_0)^{R}, \nonumber \\
\Rightarrow R & = & \frac{\log{1 - P}}{\log{1 - P_0}}
\label{eq_cum_prob_dem}
\end{eqnarray}

Hence, given that each experiment has $k_{max}$ steps and a probability of success equal to $P_c$, the total number of steps is
\begin{equation}
TS = k_{max} \frac{\log(1-P)}{\log(1-P_c)}.
\label{total_steps}
\end{equation}

In the unitary algorithm $k_{max}$ is chosen as $(\pi/4) \sqrt(N \log(N))$. In the case of $IMA$ algorithm, it is interesting to compare the results for the former choice and the optimal $k_{max}$ obtained by minimization of $TS$. 

\begin{figure}
\centering
\subfigure[$l=\sqrt{N}$ ]{
\label{e_CCM_MIcm_1} 
\includegraphics[width=0.45\textwidth]{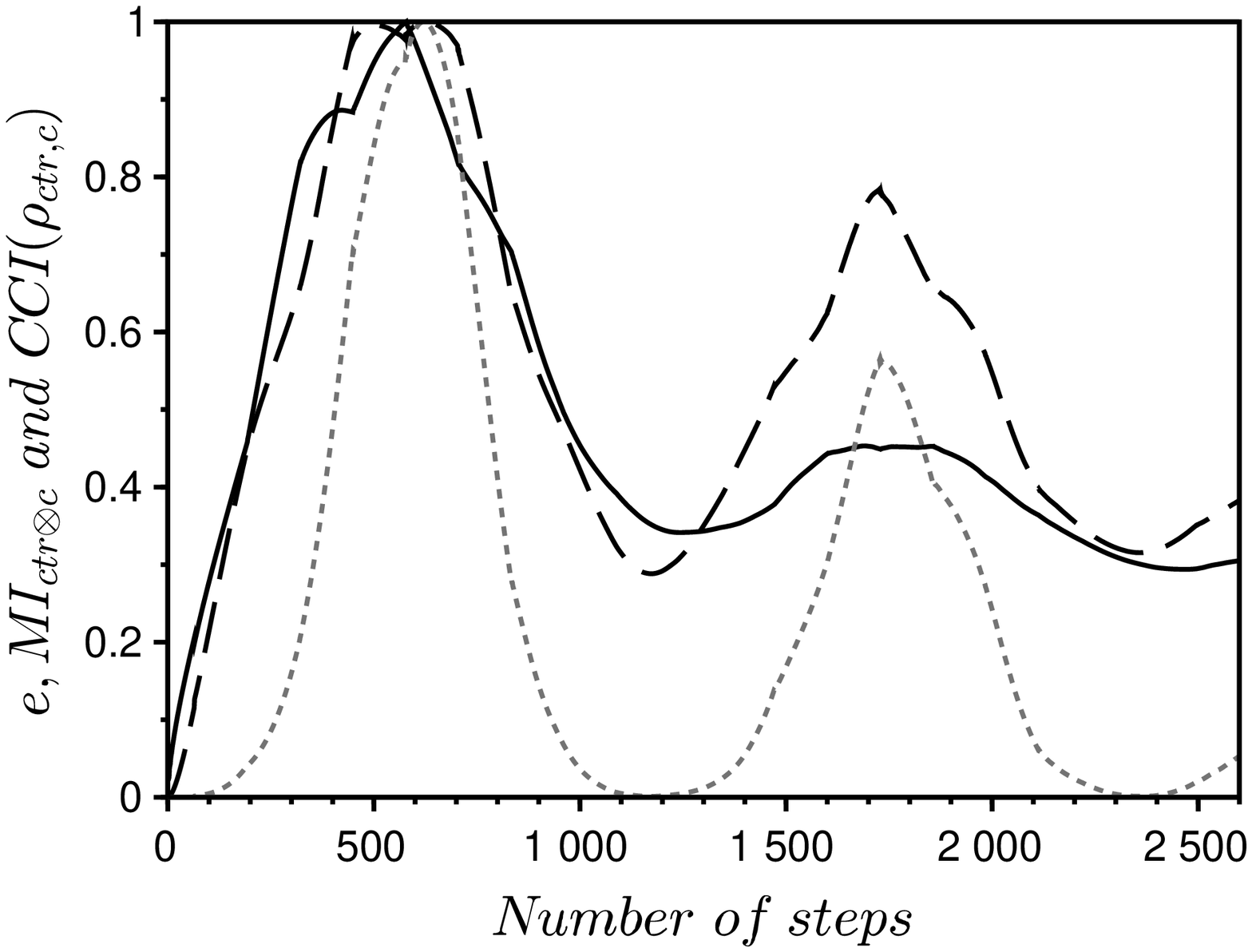}}
\hspace{0.05\textwidth}
\subfigure[$l=\sqrt{N}/2$]{
\label{e_CCM_MIcm_2} 
\includegraphics[width=0.45\textwidth]{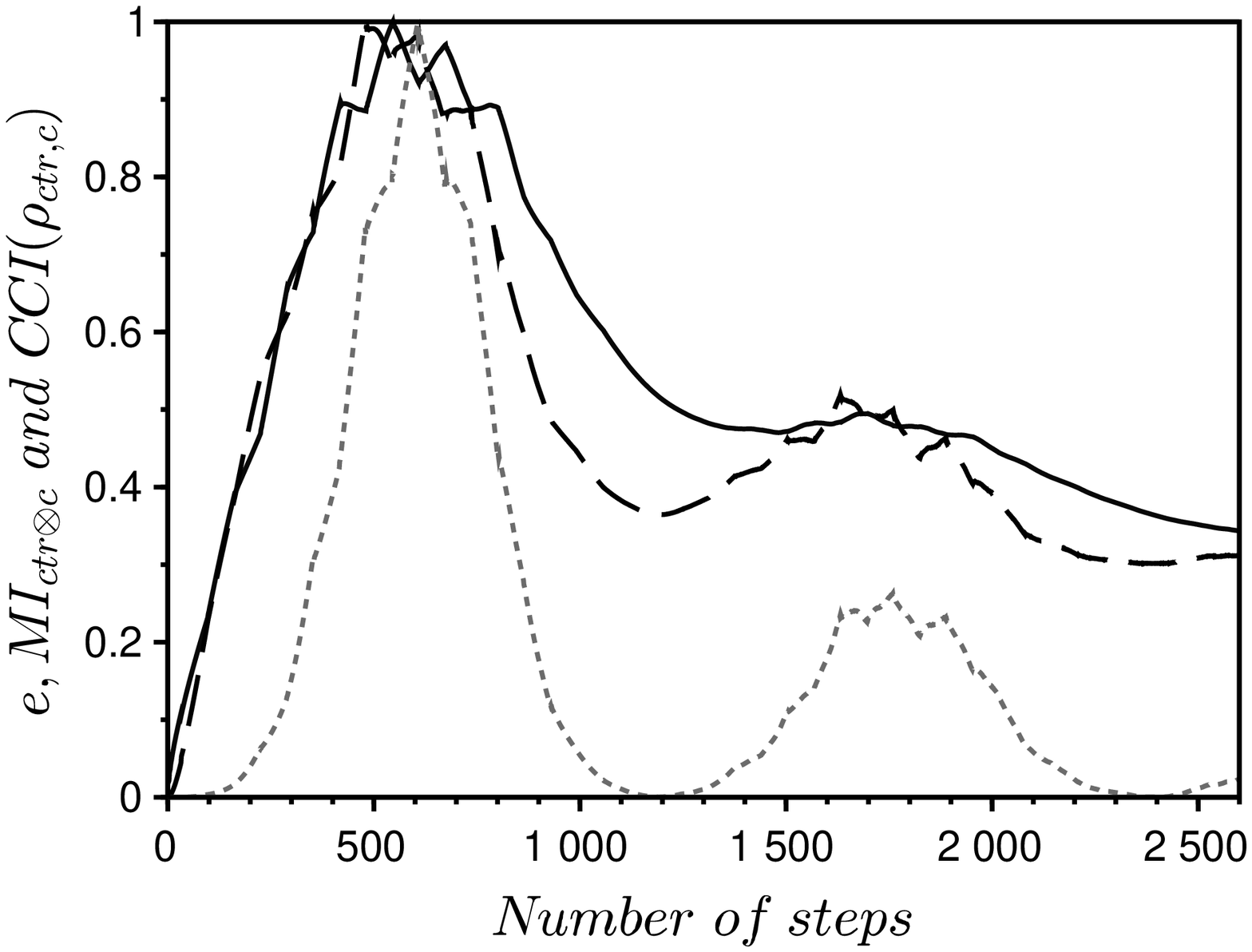}}
\subfigure[$l=\sqrt{N}/4$]{
\label{e_CCM_MIcm_4} 
\includegraphics[width=0.45\textwidth]{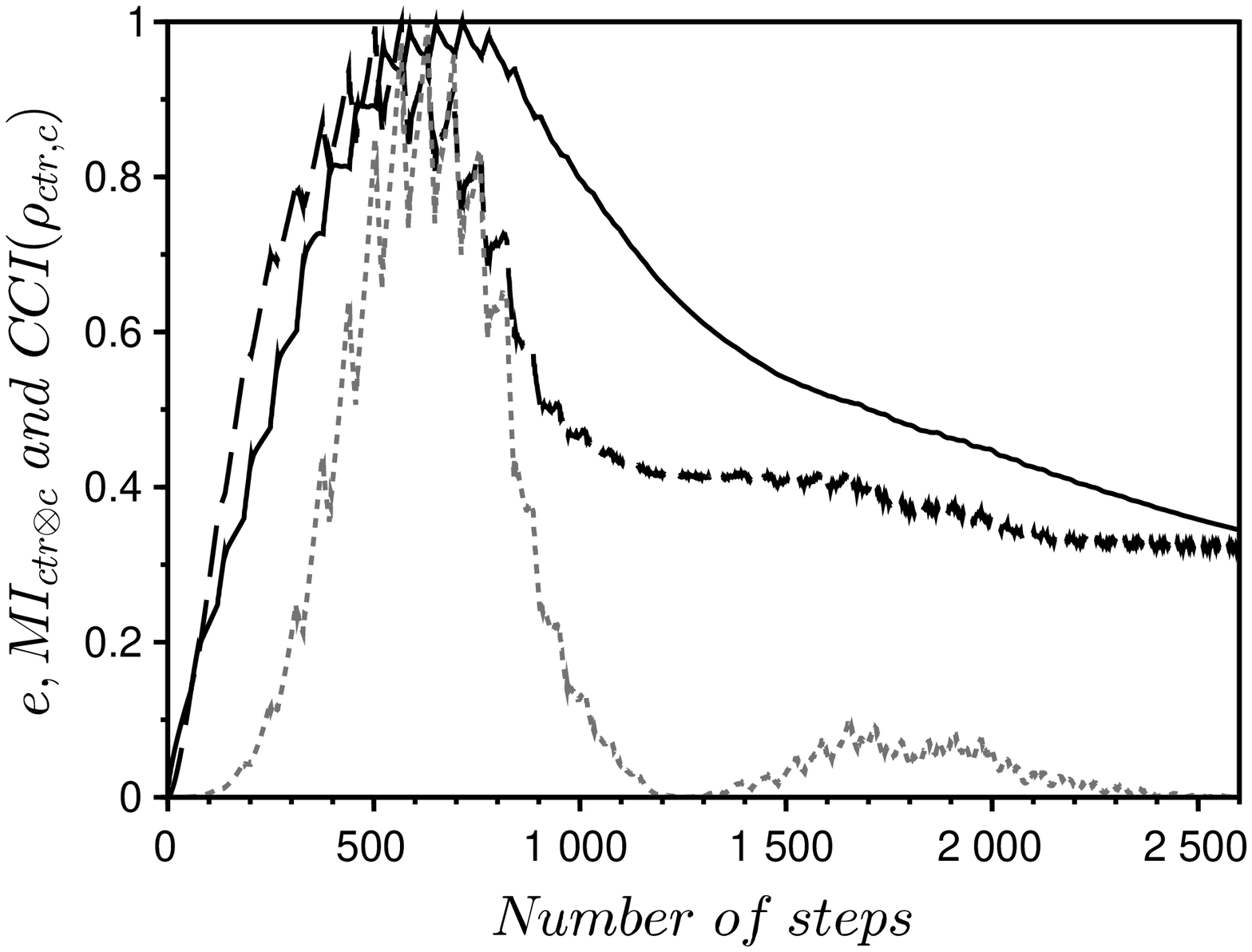}}
\hspace{0.05\textwidth}
\subfigure[$l=\sqrt{N}/8$]{
\label{e_CCM_MIcm_8} 
\includegraphics[width=0.45\textwidth]{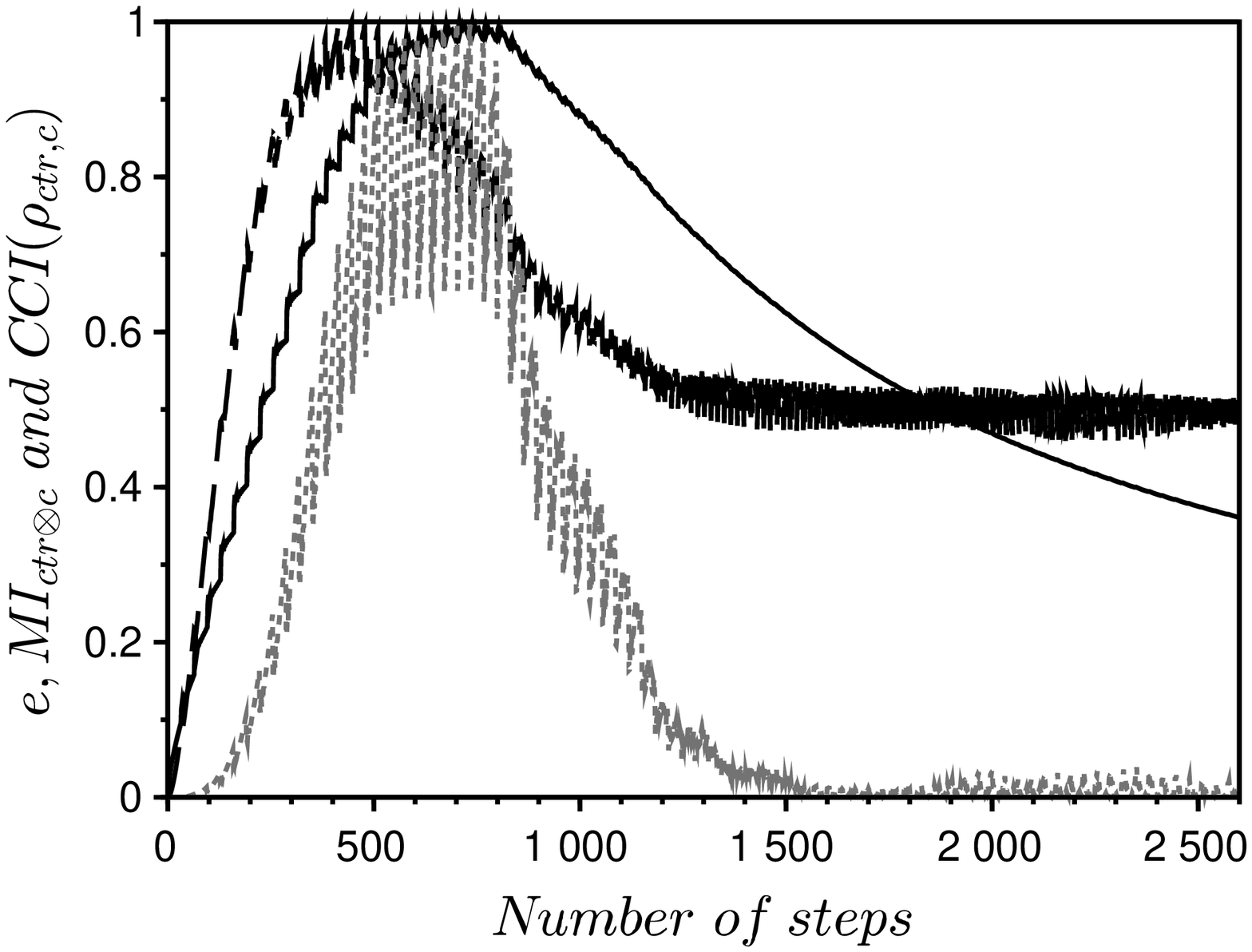}}
\caption{Correlations $CCM$(dashed black) and $MI_{ctr \otimes c}$(dotted gray), and $e(k)$, Eq. (\ref{efficiency}), versus number of steps. Curves smoothed and normalized.}
\label{cuatro_graf} 
\end{figure}

An interesting fact is that correlation in the $\mathcal{H}_{ctr \otimes c}$ subspace can be used as an indicator that approximates the point of optimal step $k_{max}$. Figure (\ref{cuatro_graf}) shows the curves $MI_{ctr \otimes c}$, $CCM ( \rho_{ctr,c})$ and $e$, where
\begin{equation}
e(k) = \frac{1}{k} \frac{\log(1-P_c)}{\log(1-P)},
\label{efficiency}
\end{equation}
being $TS = 1/e(k_{max})$. Multipartite correlations are commonly generalizations of bipartite ones, and have been used in several contexts\cite{Vedral_2002,Rulli_2011}. Multipartite correlation $CCM$\cite{Fonseca_2014} is a measure that considers, in a cumulative manner, all the bipartitions of the state space.

\begin{figure}
\centering
\subfigure[Total number of steps (expressed as $TS^{-1}$) as a function of $l$ ($m = \sqrt{N}/l$).]{
\label{fig_InvTS} 
\includegraphics[width=0.45\textwidth]{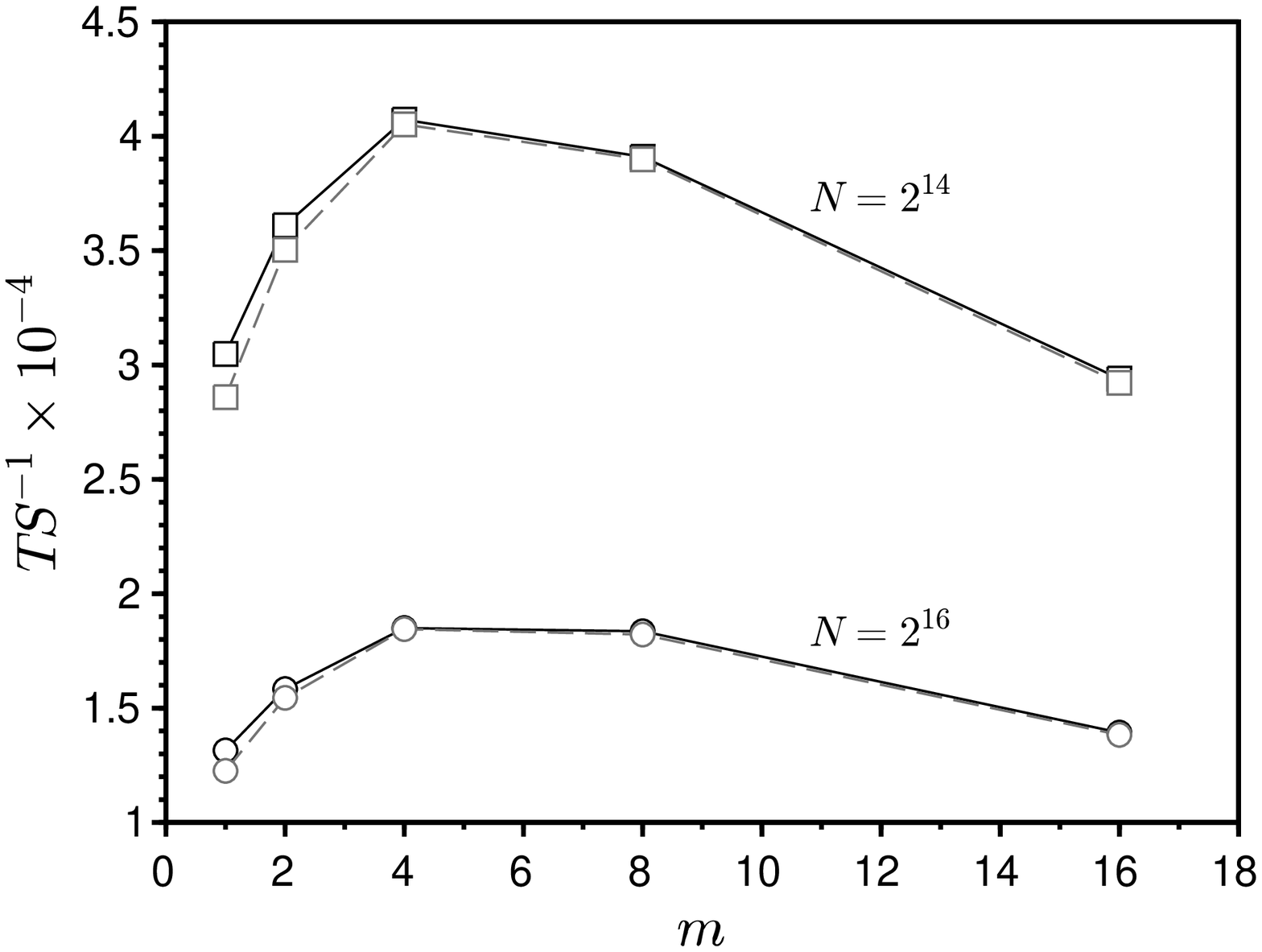}}
\hspace{0.05\textwidth}
\subfigure[Correlation $MI_{c \otimes p}$ as a function of $l$ ($m = \sqrt{N}/l$).]{
\label{fig_MImp} 
\includegraphics[width=0.45\textwidth]{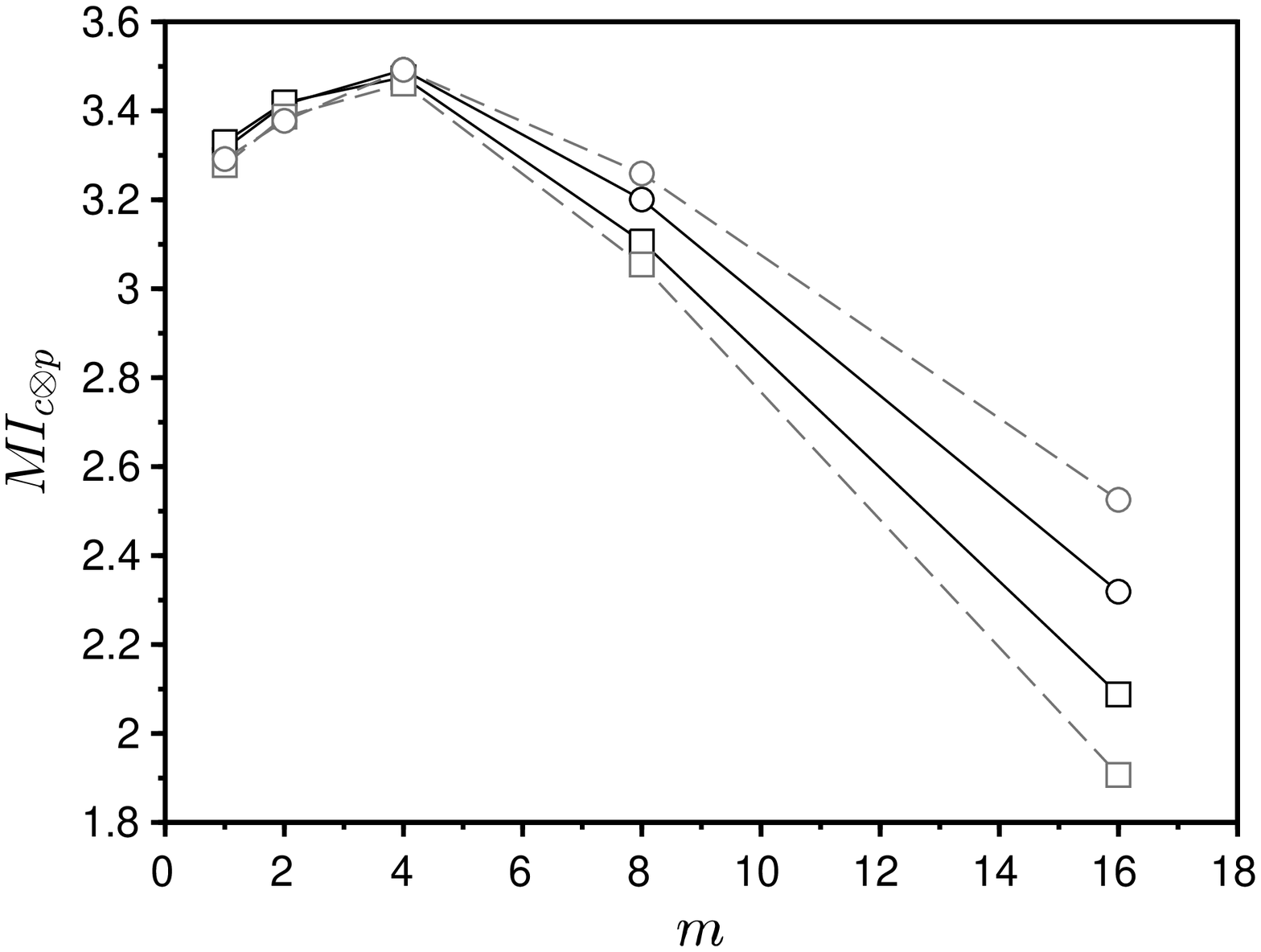}}
\caption{Comparison between $TS^{-1}$ ($e(k_{max})$) and the correlations $MI_{c \otimes p}$, for 
fixed $N=2^{14}$(square), $N=2^{16}$(circle). Both $TS$ and the correlation are smoothed and evaluated at the optimal $k_{max}$(dashed gray), and $k_{max}=\pi/4 \sqrt{N \log{N}}$ (black).}
\label{Effi_Corre} 
\end{figure}

Finally, Fig. (\ref{Effi_Corre}) shows both $e(k_{max})$ and $MI_{c \otimes p}(k_{max})$  as a function of time lapse $l$, each evaluated for a fixed $N$ ($N=2^{14}$ and $N=2^{16}$) and two different $k_{max}$ (optimal and $\pi/4 \sqrt{N \log{N}}$). All curves present a maximum for $l = \sqrt{N}/4$ ($m=4$). Interestingly, both the maximum of correlations and $e(k)$ occur for the same value of $l = \sqrt{N}/4$.

As can be observed, the results for the optimal $k_{max}$ and the standard $\pi/4 \sqrt{N \log{N}}$ are very similar, which justifies the usage of the latter.

\section{Estimating the algorithm's order}

The order of this type of search algorithm can be estimated by the number of 
steps needed to obtain a target probability of order one. $AKR$ algorithm, with quantum amplitude amplification\cite{Brassard_2002}, has order $O(\sqrt{N} \log{N})$. If classical amplitude amplification is used, the order becomes $O(\sqrt{N} \log ^{\frac{3}{2}} N)$\cite{Aaronson_2003}.

In Tulsi's algorithm, the target probability is around $\frac{1}{\cos{\delta}\sqrt{\log N}}$, resulting in an order of $O(\sqrt{ N \log N})$. For unitary $IMA_{\pi/4}$ (Tulsi with $\delta=\pi/4$) using classical amplitude amplification, the order is the same as for $AKR$, i.e. $O(\sqrt{N} \log^{3/2}{N})$.   

Motivated by the results of the previous section, where it was observed that the total number of steps $TS$ varies with time lapse $l$, in this section we estimate bounds for the order of the $IMA_{\pi/4}$ algorithm for some $l$ values.

\begin{figure}[ht!]
\centering
\subfigure[Total steps $TS$ as a function of $N$. Solid black curve with asterisk for the case $l = 1$ (measurement in all steps). Details for the others curves in Figure b.]{
\label{fig_orden_todos} 
\includegraphics[width=0.45\textwidth]{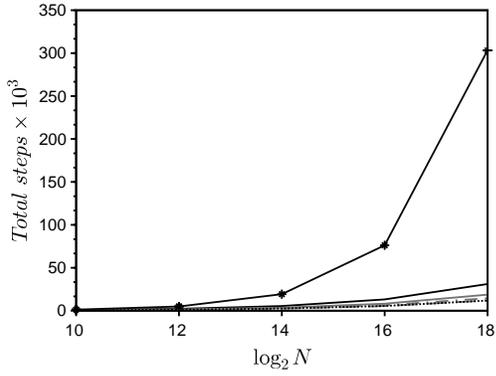}}
\hspace{0.05\textwidth}
\subfigure[Detail for the unitary case (solid black), $l = \sqrt{N}$ (solid gray), $l = \sqrt{N}/2$ (dash dotted gray), $l =  \sqrt{N}/4$ (dashed gray) and $l = \sqrt{N}/8$ (dotted black).]{
\label{fig_orden_sintodos} 
\includegraphics[width=0.45\textwidth]{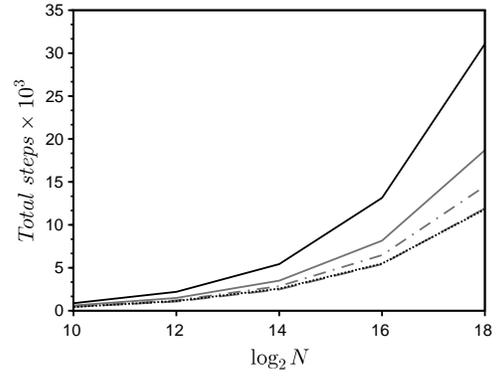}}
\subfigure[$TS$ divided by $N$. The curve for $l =1$ is the only that converges to a constant value]{
\label{fig_orden_todos_divN} 
\includegraphics[width=0.45\textwidth]{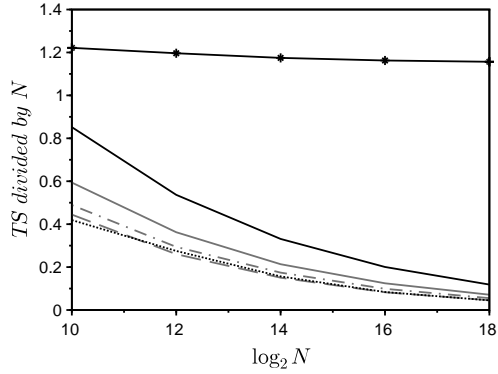}}
\hspace{0.05\textwidth}
\caption{Total steps $TS$ as a function of $N$. Curves for the unitary case and $l = 1, \sqrt{N}, \sqrt{N}/2, \sqrt{N}/4,$ and $\sqrt{N}/8$.}
\label{Estima_ordentodos} 
\end{figure}

\begin{figure}[ht!]
\centering
\subfigure[$TS$ divided by $\sqrt{N} (\log{N})^{1.5}$]{
\label{subfig_a_15} 
\includegraphics[width=0.45\textwidth]{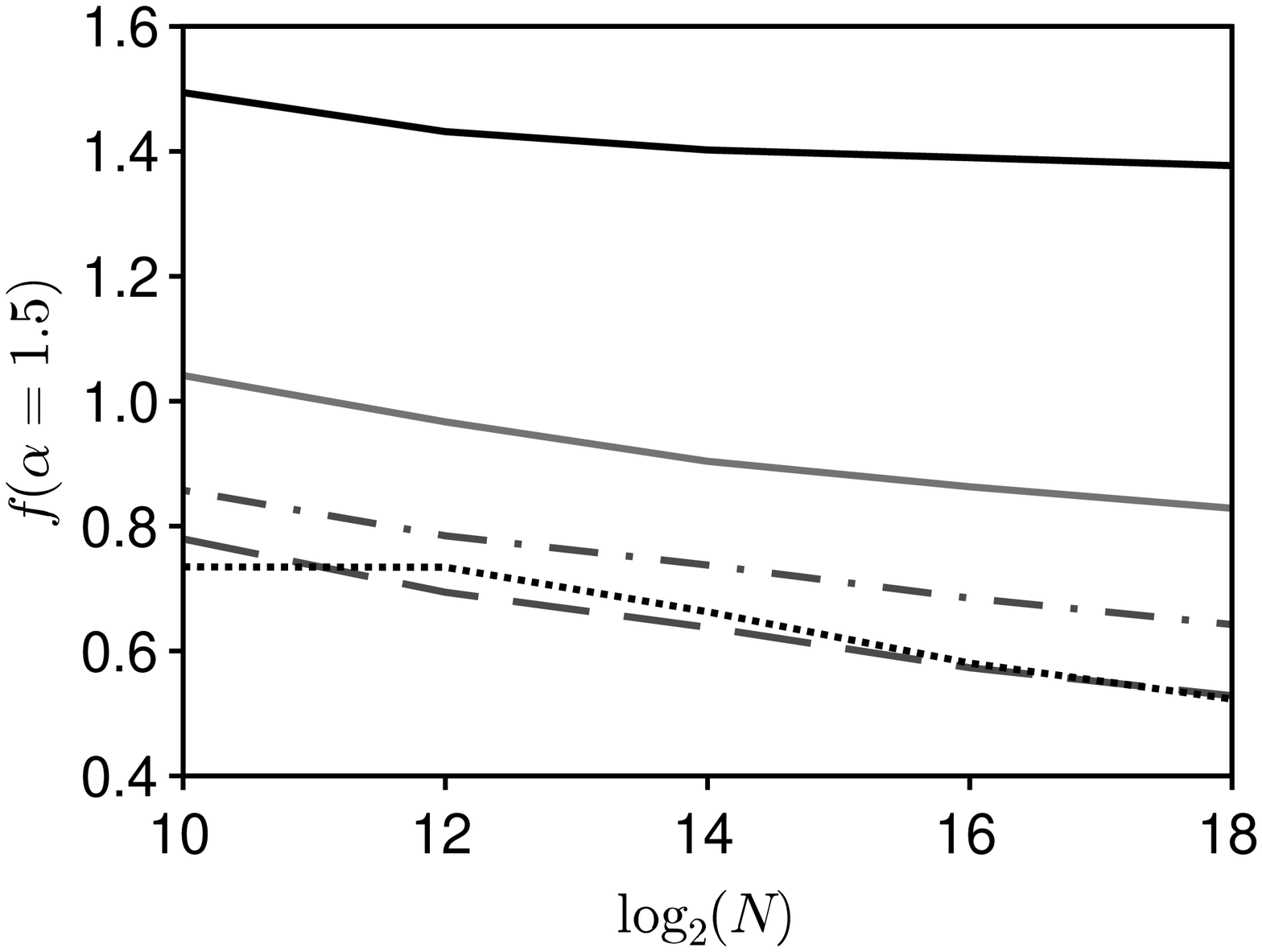}}
\hspace{0.05\textwidth}
\subfigure[$TS$ divided by $\sqrt{N} (\log{N})^{1.25}$]{
\label{subfig_a_125} 
\includegraphics[width=0.45\textwidth]{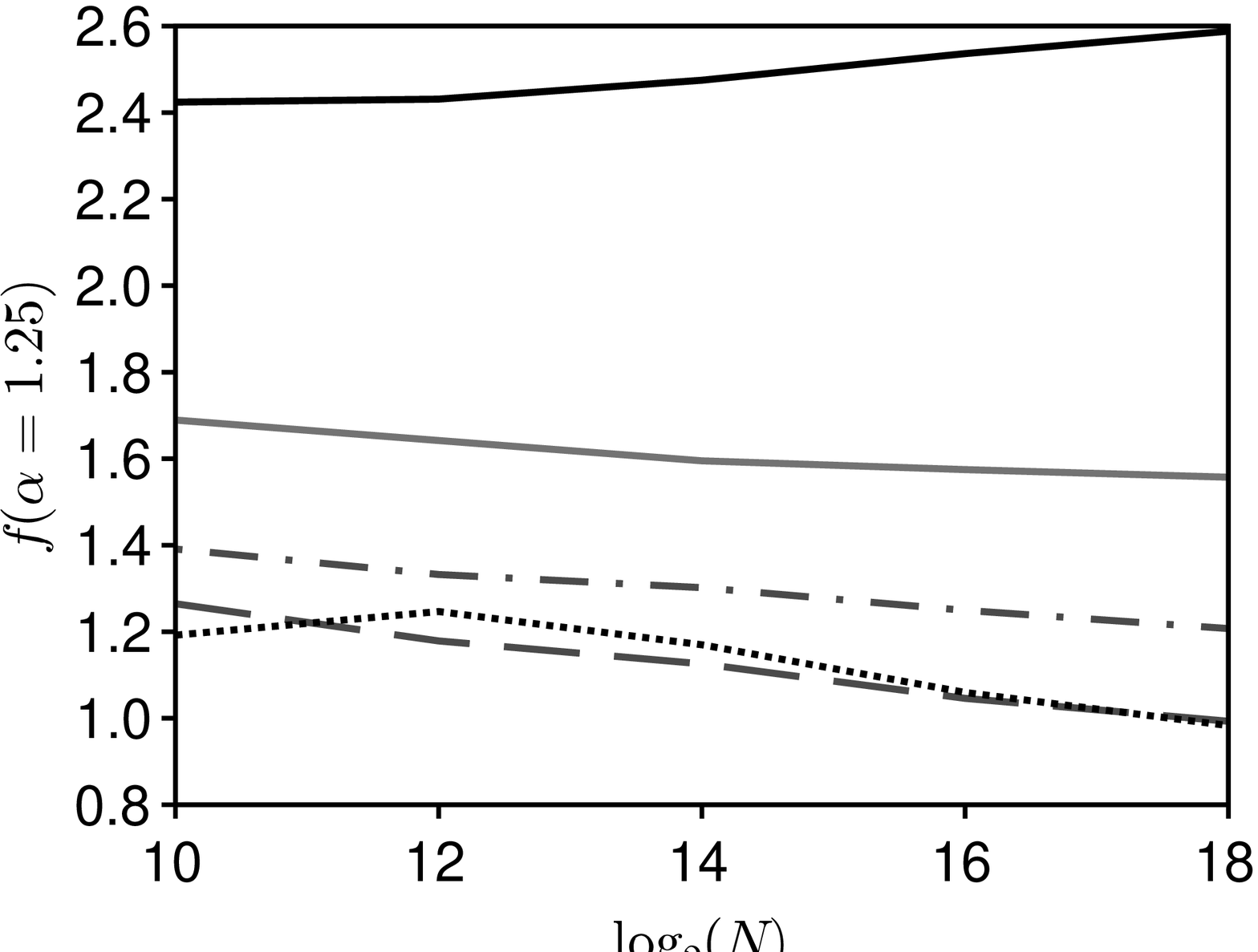}}
\subfigure[$TS$ divided by $\sqrt{N} (\log{N})^{0.9}$]{
\label{subfig_a_09} 
\includegraphics[width=0.45\textwidth]{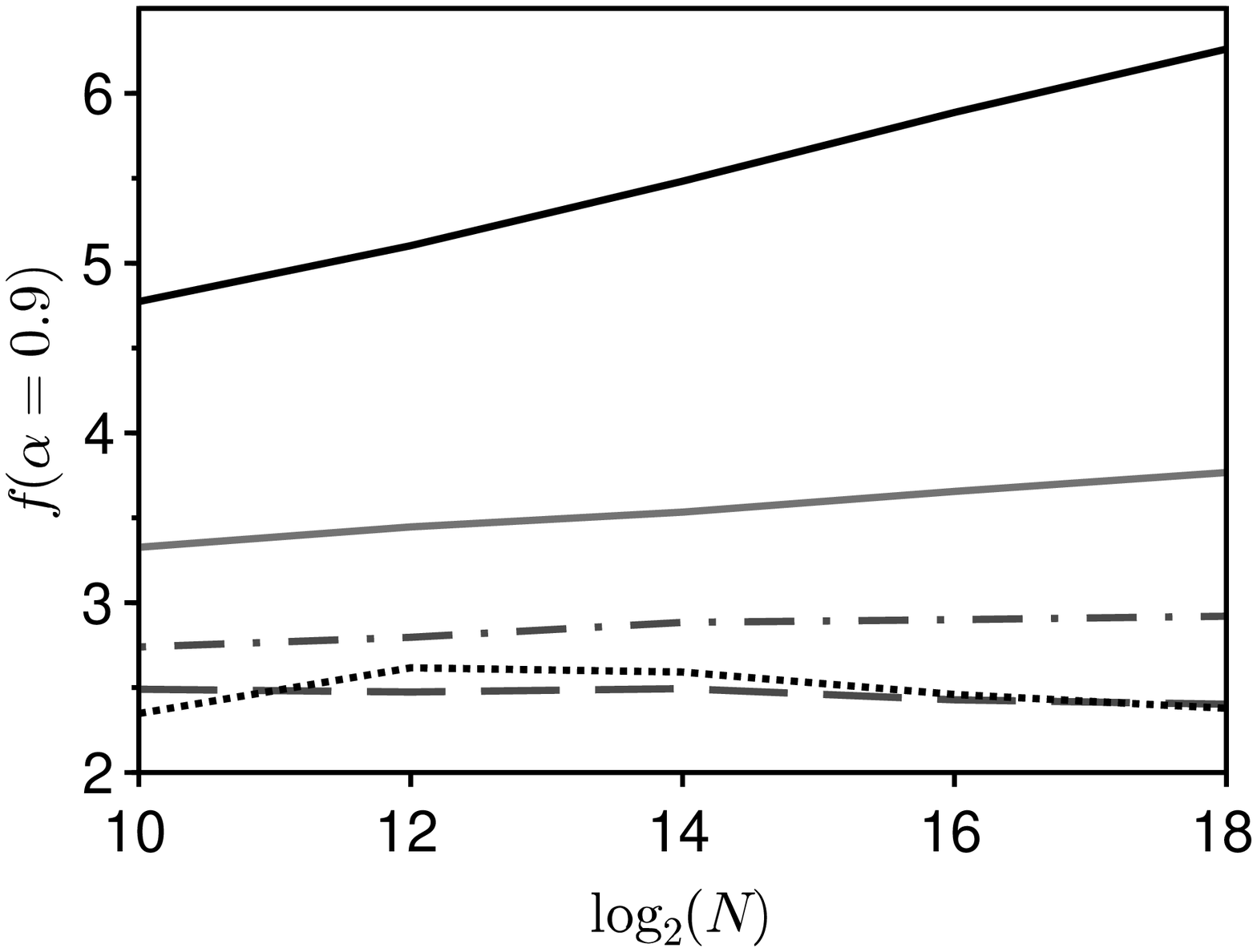}}
\hspace{0.05\textwidth}
\subfigure[$TS$ divided by $\sqrt{N} (\log{N})^{0.6}$]{
\label{subfig_a_06} 
\includegraphics[width=0.45\textwidth]{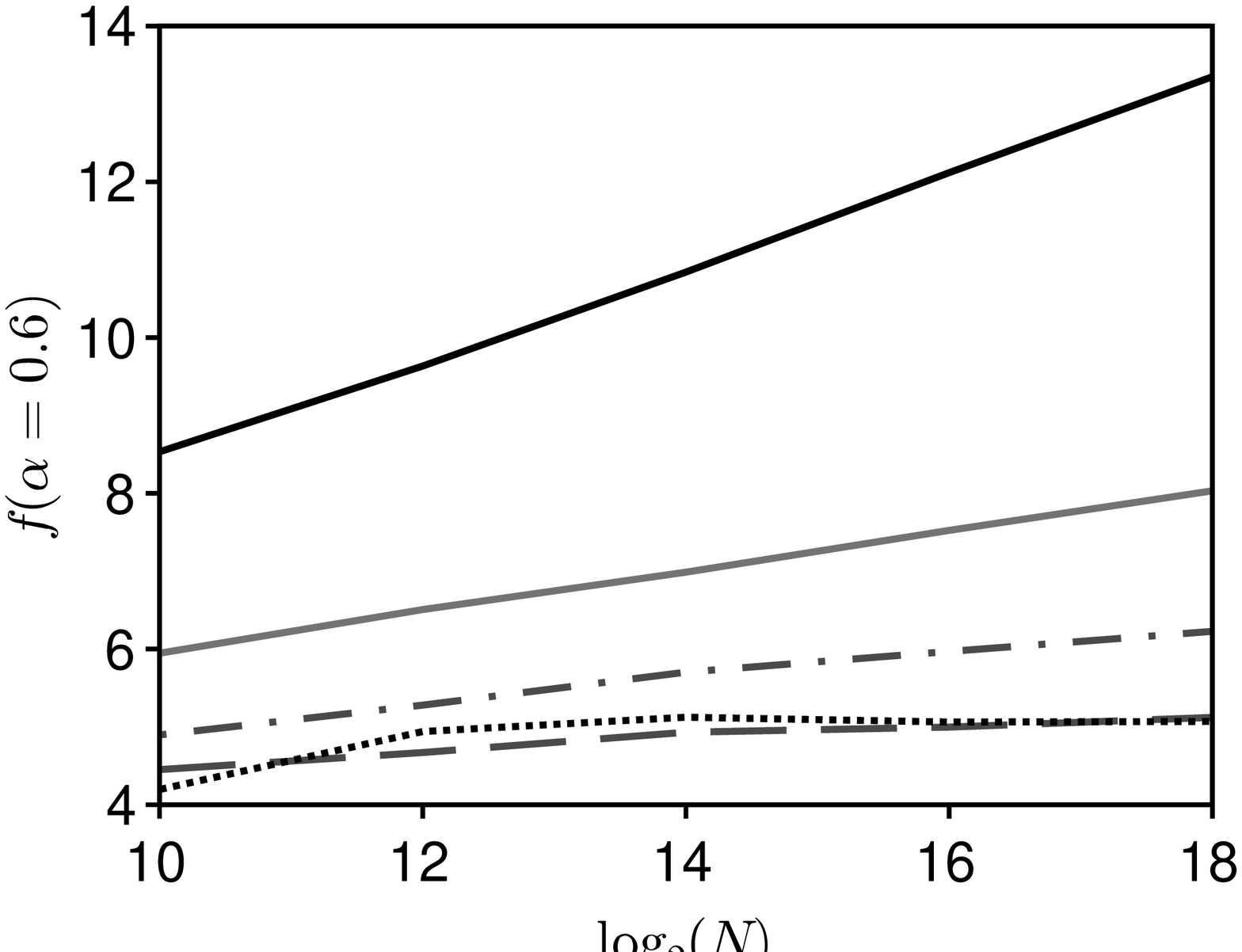}}
\caption{Total steps $TS$ divided by fitting curve $ \left( f(\alpha) =  \frac{TS}{\sqrt{N} (\log{N})^\alpha} \right)$. Unitary case (solid black), $l = \sqrt{N}$ (solid gray), $l = \sqrt{N}/2$ (dash dotted gray), $l =  \sqrt{N}/4$ (dashed gray) and $l = \sqrt{N}/8$ (dotted black).}
\label{Estima_orden} 
\end{figure}

Figure (\ref{Estima_ordentodos}) shows the total number of steps $TS$ as a function of $N$
, for some $l$ values. As can be seen in Figure (\ref{Estima_ordentodos}.c), when divided by $N$, the curve corresponding to the $l=1$ case is the only curve that converges to a constant value. In this case the algorithm has the same order as the classical brute-force search algorithm. Other curves show a better order than classical. Due to computational limitations to obtain results for large values of $N$, it is very 
hard to perform good nonlinear regressions to fit the order. Instead, we estimate ranges 
for the order depending on $l$, dividing the curves in Figure (\ref{Estima_ordentodos}.b) by functions of type
\begin{equation}
\beta \sqrt{N} (\log{N})^\alpha,
\label{Curva_orden}
\end{equation}
for some $\alpha$. Figure (\ref{Estima_orden}) shows the results for $\alpha$ equal to: $1.5$, $1.25$, $0.9$ and $0.6$.
 
It can be observed that:
\begin{itemize}
\item The curve corresponding to the unitary algorithm tends to be constant when divided by Eq. (\ref{Curva_orden}) with $\alpha = 1.5$ (Figure (\ref{Estima_orden}).(a)), as expected.  
\item For $l = \sqrt{N}$ and $\sqrt{N}/2$ the order is in the range $\alpha \in \left [0.9, 1.25 \right]$, as can be appreciated in Figures (\ref{Estima_orden}).(c) and (\ref{Estima_orden}).(b).
\item In the cases $l = \sqrt{N}/4$ and $\sqrt{N}/8$, Figures (\ref{Estima_orden}).(c) and (\ref{Estima_orden}).(d) shows that the order is in the range $\alpha \in \left [0.6, 0.9 \right]$.
\end{itemize}

\section{Conclusions}
In this paper a modified Tulsi's algorithm with intermediate 
partial measurements of the control qubit($IMA_{\pi/4}$), is presented. 

The target probability $P_t$, and some correlations (Section \ref{subsec_IMA_prob_cor_Pt}), behave similarly to energy of a damped harmonic oscillator, where time lapse $l$ works as a decoherence parameter, going from quantum to classical. 

The performance of the algorithm also has a strong relation with some correlations. 
The maxima in the $MI_{ctr \otimes c}$ and $CCM$ curves indicate the optimal step to stop the algorithm, when classical amplification is considered.
What is more, as it can be observed from Figures (\ref{cuatro_graf}) and (\ref{Effi_Corre}), when these maxima have better coincidence ($l = \sqrt{N}/4$) the algorithm $IMA_{\pi/4}$ has minimal total steps $TS$.

For some values of $l$, the order estimated shows an improvement with respect to the 
unitary case. However, a numerical approach to find the order is limited by computational power. This fact motivates, as a future work, the search of analytical approaches to this problem. 

This study, with partial intermediate measurements in quantum search algorithms, is a start point to analyze possible improvements of other quantum algorithms using one, or several, control qubits.

\section*{Acknowledgments}
We thank Marcelo Terra-Cunha, and Laura Diaz Ernesto, for discussions and comments about the article.

\bibliographystyle{unsrt}
\bibliography{IMQW_2014}

\end{document}